\documentclass[aps,prl,twocolumn,showpacs,superscriptaddress,groupedaddress]{revtex4} 
\pdfoutput=1
\usepackage[version=3]{mhchem} 
\usepackage[dvips]{epsfig}
\usepackage{graphicx}  
\usepackage{dcolumn}   
\usepackage{bm}        
\usepackage{amssymb}   

\begin{document}

\author{Paul E. Barclay}
\altaffiliation{ Institute for Quantum Information Science, University of Calgary, Calgary, AB, T2N 1N3, Canada}
\altaffiliation{ NRC-National Institute for Nanotechnology, 11421 Saskatchewan Drive NW, Edmonton, AB T6G 2M9, Canada}
\email{pbarclay@ucalgary.ca}
\author{Kai-Mei C. Fu}
\altaffiliation{ University of Washington, 4014 University Way NE, Seattle, WA 98105 USA}
\author{Charles Santori}
\affiliation{Hewlett-Packard Laboratories, 1501 Page Mill Road, Palo Alto, CA 94304 USA}
\author{Andrei Faraon}
\affiliation{Hewlett-Packard Laboratories, 1501 Page Mill Road, Palo Alto, CA 94304 USA}
\author{Raymond G. Beausoleil}
\affiliation{Hewlett-Packard Laboratories, 1501 Page Mill Road, Palo Alto, CA 94304 USA}

\title{Hybrid nanocavities for resonant enhancement of color center emission in diamond}

\begin{abstract}
Resonantly enhanced emission from the zero phonon line of a diamond nitrogen-vacancy (NV) center in single crystal diamond is demonstrated experimentally using a hybrid whispering gallery mode nanocavity.  A 900~nm diameter ring nanocavity formed from gallium phosphide, whose sidewalls extend into a diamond substrate, is tuned onto resonance at low-temperature with the zero phonon line of a negatively charged NV center implanted near the diamond surface.  When the nanocavity is on resonance, the zero phonon line intensity is enhanced by approximately an order of magnitude, and the spontaneous emission lifetime of the NV is reduced as much as $18\%$, corresponding to a 6.3X enhancement of emission in the zero photon line.
\end{abstract}

\maketitle

\noindent The diamond nitrogen-vacancy (NV) center is an optically active impurity which combines many of the desirable properties of quantum dots and laser trapped atoms. Optical transitions of diamond NV centers can display low inhomogeneous broadening, and have been used to generate single photons \cite{ref:kurtsiefer2000sss},  manipulate single electron spins \cite{ref:jelezko2004oco}, and control nearby nuclear spin impurities \cite{ref:jelezko2004ocs,ref:childress2006cdc,ref:gurudevdutt2007qrb}. Remarkably, room temperature electron spin coherence times of NVs can exceed a millisecond \cite{ref:balasubramanian2009usc}.   These properties make NVs a promising qubit for proposed quantum networks \cite{ref:kimble2008qi}, and an attractive system for applications such as magnetometry \cite{ref:maze2008nms} and low power optical switching \cite{ref:mabuchi2009cqm}.  An outstanding challenge to using NV centers as qubits in quantum information processing applications is creating a platform which mediates interactions between them.  A promising approach to this problem is to create an on-chip quantum network, in which NVs interact optically via nanophotonic interconnects \cite{ref:benjamin2009pfm}.  Coupling NVs to optical cavities plays a crucial role in this implementation, by enhancing the NV emission into a well defined optical mode, which can be efficiently coupled to waveguides and routed on-chip.  Cavity enhancement of emission is particularly important for NV centers, as it provides a means for increasing the relative brightness of narrowband zero phonon line (ZPL) emission relative to broadband phonon assisted emission.  Selection of emission into the ZPL is necessary for protocols involving coherent interactions or indistinguishable photons.

\begin{figure}[b]
\begin{center}
  \epsfig{figure=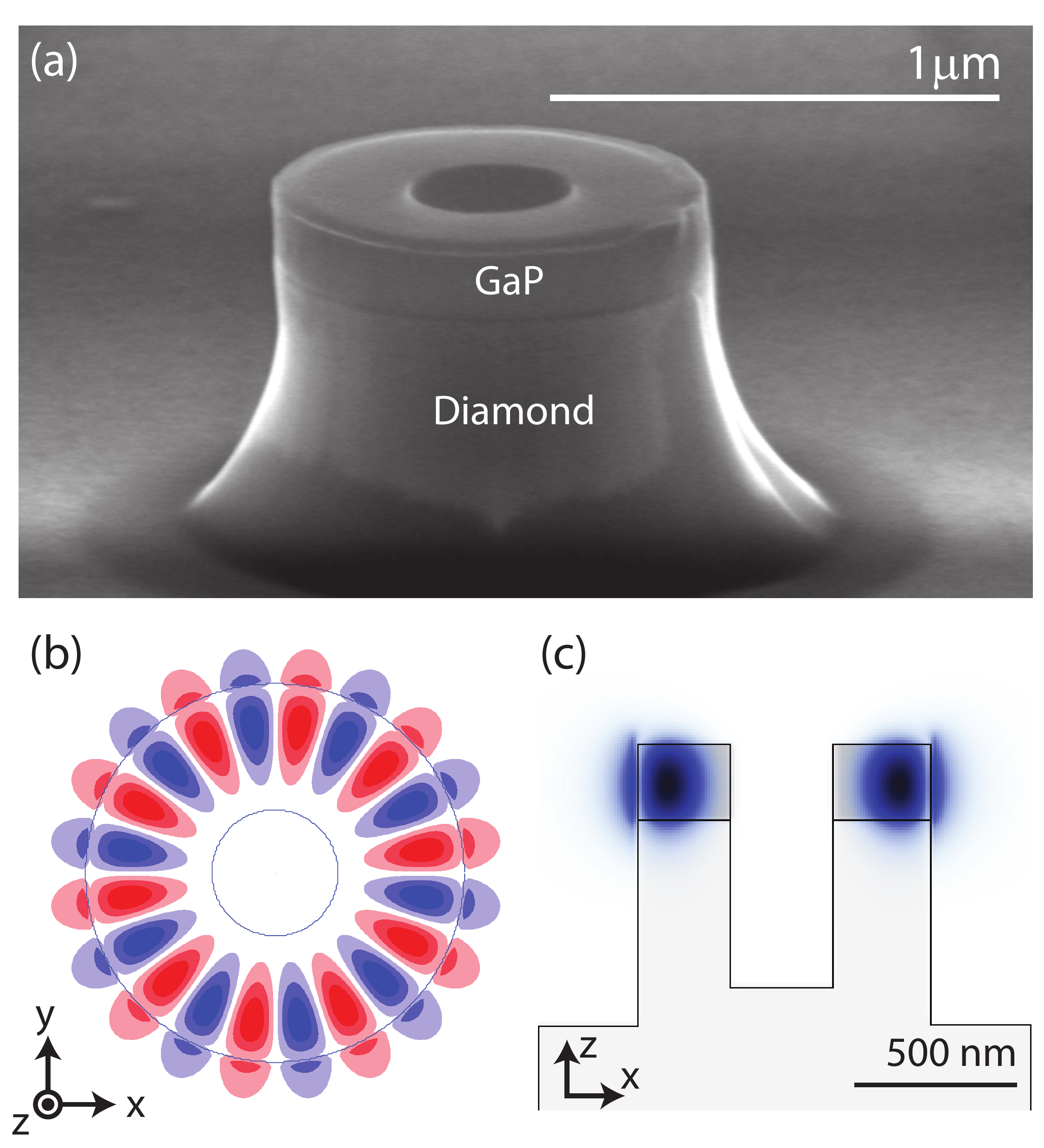, width=1\linewidth}
  \caption{(a) SEM image of a hybrid GaP-diamond whispering gallery mode nanocavity. (b) Top-view and (c) cross-section of the dominant electric field component, $E_r$, of the lowest order TE-like standing wave mode supported by the device in (a), with resonance wavelength $\lambda_o \sim 637~\text{nm}$. The fields are calculated using finite difference time domain simulations.  The field in (b) is plotted in the $x-y$ center-plane of the GaP layer.}
  \label{fig:sem}
\end{center}
\end{figure}

\begin{figure*}[t]
\begin{center}
  \epsfig{figure=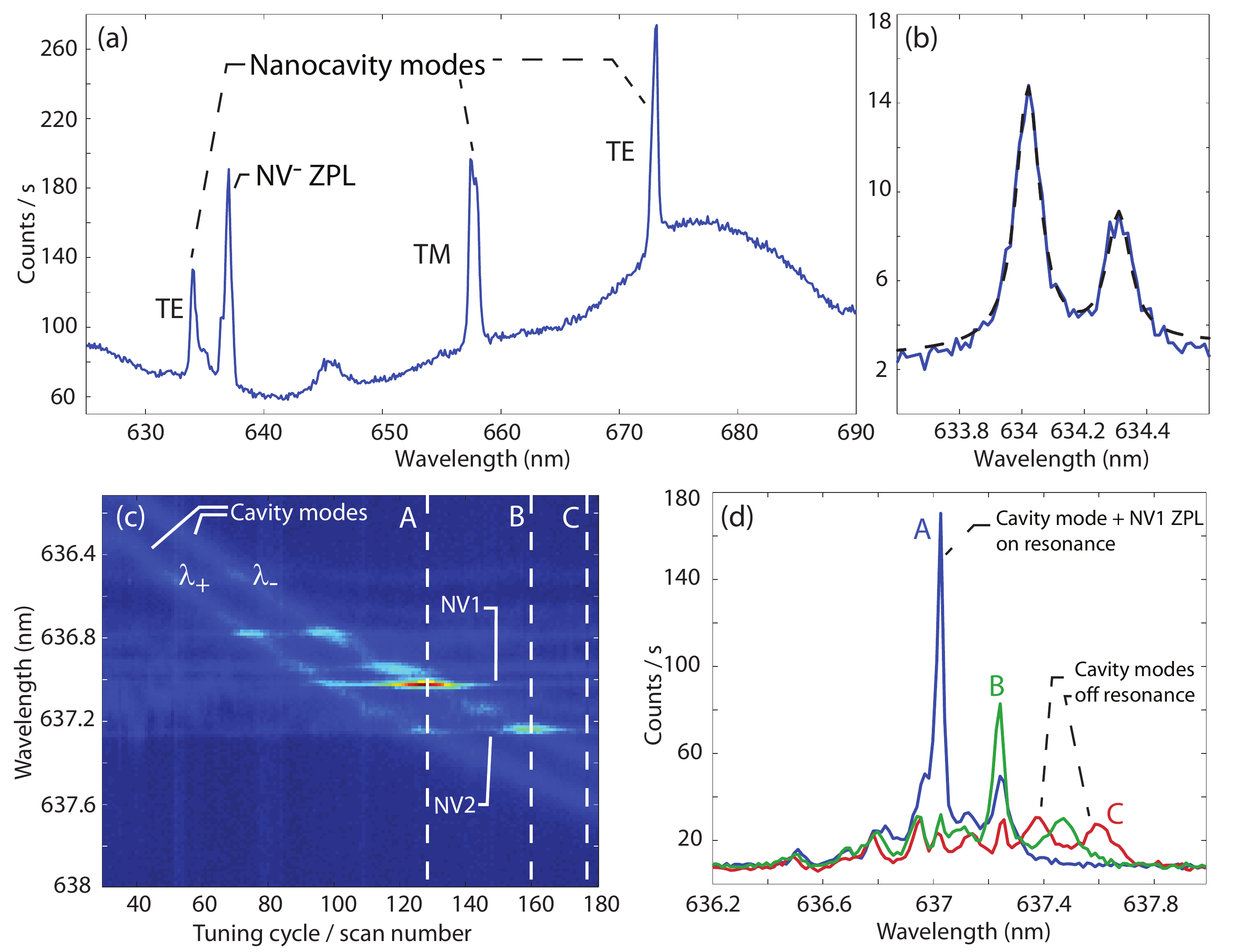, width=0.8\linewidth}
  \caption{ (a)  Spectrum of the nanocavity studied in (c) and (d) prior to tuning the nanocavity modes. (b) High-resolution PL spectrum of the nanocavity mode in (a) closest to the NV$^-$ ZPL. The dashed line is a fit to the data of two incoherently added Lorentizian lineshapes.  (c) Nanocavity PL spectrum as a function of cavity tuning cycle.  Each tuning cycle corresponds to releasing a fixed volume of Xe gas into the cryostat. (d) Nanocavity PL spectra when the nanocavity is on resonance with the ZPL of NV1 and NV2 (spectra A and B, respectively), and off resonance  from any NV ZPL (spectra C).  Spectra A, B, and C are measured after the tuning cycles indicated by the dashed lines in (c).}
  \label{fig:tuning}
\end{center}
\end{figure*}

\begin{figure*}[htb]
\begin{center}
  \epsfig{figure=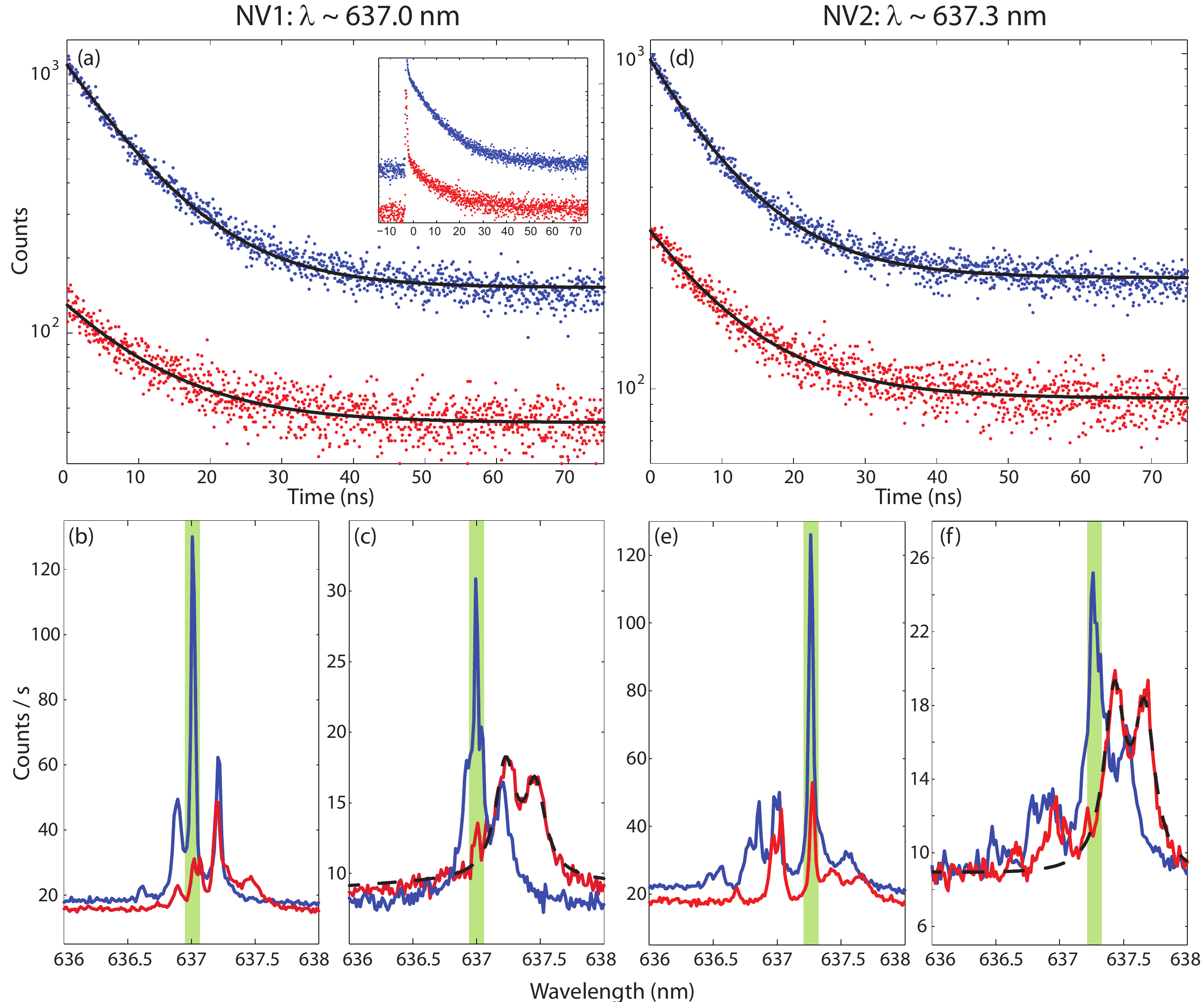, width=0.8\linewidth}
  \caption{(a) Time resolved photoluminescence of NV1 excited with a pulsed green source, when the NV1 ZPL is on (blue) and off (red) resonance with the nanocavity $\lambda_-$ mode.  The time origin is chosen $\sim 3.0\text{ns}$ after the excitation pulse peak (see inset) so that fast decaying nanocavity background does not influence the fits (solid lines).  Nanocavity spectra under (b) CW, and (c) pulsed 532~nm excitation, when NV1 ZPL is on (blue) and off (red) resonance with the nanocavity mode.  The green shaded regions indicate the monochromator spectral window used for the lifetime measurements in (a). The dashed line in (c) is a fit to the data consisting of two incoherently superimposed Lorentzian lineshapes. (d,e,f) Data analogous to (a,b,c), with the nanocavity $\lambda_-$ mode tuned on and off resonance with the NV2 ZPL. }
  \label{fig:lifetime}
\end{center}
\end{figure*}

Recent efforts to efficiently couple NVs in nanocrystalline diamond to nanophotonic structures \cite{ref:barclay2008cie, ref:englund2010dcs, ref:wolters2010ezp, ref:vandersar2011dnc} have been limited by poor NV optical properties in nanocrystals compared to those found in  single crystal diamond.   Progress towards fabricating nanophotonic devices directly from single crystal diamond has recently made important progress \cite{ref:faraon2011rez}, limited primarily by fabrication difficulties related to creating thin films of single crystal diamond necessary for optical confinement in three dimensions.   An alternative  approach, which leverages existing semiconductor processing technology, is to create photonic structures from hybrid material systems in which a thin waveguiding layer is bonded to the surface of a single crystal diamond substrate.  Light localized within waveguides  and microcavities lithographically defined in the waveguiding layer can interact evanescently with NVs positioned near the surface of the diamond substrate, making this a natural system for coupling to arrays of NV implanted near surfaces, such as those studied in Ref.\ \cite{ref:toyli2010csn}. The hybrid approach can take advantage of properties of the waveguiding material which are not available in all-diamond systems, for example nonlinear or optoelectronic response useful for integrated optical modulation \cite{ref:yariv2006poe}. Hybrid semiconductor-diamond devices were used in Refs. \cite{ref:barclay2009cbm,ref:fu2008cnv} to demonstrate evanescent coupling between ensembles of NVs and micron-scale photonic waveguides and cavities.  Here, we demonstrate optical coupling between a nanoscale hybrid optical cavity and a single diamond NV center, and measure resonant Purcell enhanced spontaneous emission into the ZPL.

The nanocavities studied here were realized from a hybrid geometry consisting of a gallium phosphide (GaP, 250~nm thickness, $n_\text{GaP} \sim 3.3$) whispering gallery mode nanocavity supported by a single crystal diamond substrate.  A typical device is shown in Fig.\ 1(a). They were fabricated following the process in Ref. \cite{ref:barclay2009cbm}.  The diamond substrate consists of a CVD grown electronic grade single crystal diamond sample (Element Six) subjected to ion implantation (N$^+$ 10keV, $1\times10^{10}$cm$^{-2}$)  and annealing (900$^\text{o}$C in H$_2$/Ar) to create NVs close to the diamond surface. A subsequent oxygen anneal step maximized the NV$^-$/NV$^0$ ratio near the surface \cite{ref:fu2010cnn}.   The nanocavity diameter, $d \sim 900~\text{nm}$, is 5X smaller than in previous work \cite{ref:barclay2009cbm}.  To enhance optical confinement, the GaP sidewalls were extended $\sim 600~\text{nm}$ into the diamond using an oxygen plasma etch, decreasing the effective refractive index of the underlying nanocavity substrate.

This structure supports whispering gallery modes whose field is primarily confined inside the GaP  and interacts evanescently with NVs near the diamond surface.  Figures \ref{fig:sem}(b) and (c) show the simulated field profile of the whispering gallery mode supported by this structure with azimuthal mode index $m=9$, fundamental radial order ($p=0$), and TE-like polarization (dominant electric field component radially polarized) .  The field profile was calculated using a finite difference time domain simulation (FDTD) \cite{ref:oskooi2010mff}, and has a resonance wavelength close to $\lambda_\text{ZPL} \sim 637~\text{nm}$ of the NV$^-$ ZPL, mode volume $V \sim 3.0(\lambda/n_\text{GaP})^3$, defined by the peak electric field energy density, and a maximum intensity in the diamond of $\eta_\text{dia} \sim 0.11$ of the peak intensity in the GaP.  The theoretical radiation limited quality factor, $Q_\text{rad}$, for these structures exceeds $10^6$; in practice fabrications imperfections and material absorption will limit $Q$ below $Q_\text{rad}$ \cite{ref:barclay2009cbm}. 

The fraction of spontaneous emission radiated from an  evanescently coupled NV into the nanocavity mode described above can be predicted by calculating the Purcell enhancement to the zero phonon emission at $637$~nm.  In bulk, an NV$^-$ radiates a fraction $\zeta_\text{ZPL} \sim 3\%$ of its emission into the ZPL, estimated from our measurements of area under the ZPL in the spontaneous emission spectrum from an ensemble of NVs.  In the presence of a resonant nanocavity, emission into the ZPL is enhanced by a factor $F_\text{ZPL}$ due to the Purcell effect \cite{ref:purcell1946sep,ref:santori2010nqo}:
\begin{equation}\label{eq:F}
F_\text{ZPL} = \frac{3}{4\pi^2}\frac{n_\text{o}}{n_\text{d}}\left(\frac{\lambda_\text{ZPL}}{n_\text{GaP}}\right)^3\frac{Q}{V}\left|\frac{\mathbf{\mu}\cdot \mathbf{E}(\mathbf{r}_\text{NV})}{\mathbf{E}_\text{o}}\right|^2,
\end{equation}
where $\mathbf{E}(\mathbf{r}_\text{NV})$ is the electric field strength at the NV, $\mathbf{E}_\text{o}$ is the peak nanocavity field strength, $\mu$ is a unit vector describing the NV dipole orientation, 
$n_\text{o}$ is the refractive index at the peak field location, and $n_\text{d}$ is the refractive index of diamond.   The total cavity enhanced NV$^-$ spontaneous emission rate is then given by
\begin{equation}\label{eq:gamma}
\gamma_\text{c} \sim \gamma_\text{o}(1 + F_\text{ZPL}\zeta_\text{ZPL}),
\end{equation}
where $\gamma_\text{o}$ is the bulk NV$^-$ spontaneous emission rate.  Equation \ref{eq:gamma} assumes that the NV-cavity system is in the ``bad cavity'' limit, and neglects the modification of off resonance NV emission by the nanocavity.

The nanocavity optical properties were studied by exciting the device with a 532~nm source and measuring the resulting photoluminescence (PL).  A scanning confocal microscope (0.6 NA objective) was used to excite and collect PL from a sub-micron diameter spot on the nanocavity.  Collected light was directed to a spectrometer or a time resolved photon counting module.  For the implantation dose used here, the NV density is such that  a small number ($\sim$ 1 - 10) of NVs were typically excited by the excitation spot.   All of the measurements were performed with the sample mounted in a liquid helium flow cryostat and cooled to 6K.  Figure \ref{fig:tuning}(a) shows a broad wavelength, low resolution (0.14~nm), spectrum of the nanocavity studied here when it was excited with a CW 532~nm source ($\sim 1$~mW). Emission from the 637~nm  ZPL of negatively charged NVs is clearly visible.  Peaks in the emission corresponding to PL coupled to nanocavity modes are also evident, including a TE mode  blue-detuned 3~nm from the NV$^-$ ZPL.  The modal polarization labels indicated in Fig.\ \ref{fig:tuning}(a)  were determined by comparing the resonance spacing with FDTD predicted values, as in Ref.\ \cite{ref:barclay2009cbm}, and by comparing their relative tuning rates in the measurements described below.

Figure \ref{fig:tuning}(b) shows a high resolution ($\sim 0.02$~nm) spectrum of PL from the nanocavity mode closest to the NV$^-$ ZPL.  This reveals that the nanocavity mode has a doublet structure consisting of two peaks at wavelengths $\lambda_\pm=\lambda_o\pm\Delta\lambda/2$ with full-width at half-max $\delta\sim 94$~pm ($Q_{\pm} \sim 6800$).  $\Delta\lambda=0.27$~nm is the doublet splitting, and  $\delta$ was determined by fitting the data with two incoherently superimposed Lorentzian lineshapes, as shown in Fig.\ \ref{fig:tuning}(b).   The doublet structure results from nanocavity surface roughness and imperfections which couple the nominally degenerate clockwise and counterclockwise circulating whispering gallery modes, creating non-degenerate standing wave modes \cite{ref:borselli2005brs}.

As described by Eq.\ \ref{eq:F}, when a nanocavity mode is resonant with the NV  ZPL, the NV spontaneous emission rate can be enhanced through the Purcell effect. We demonstrate this here by tuning the nanocavity in  Fig.\ \ref{fig:tuning}(a) through resonance with the NV$^-$ ZPL, and showing that the nanocavity  significantly enhances the ZPL intensity of  coupled NVs.  The nanocavity resonances are tuned by injecting Xe gas into the cryostat \cite{ref:mosor2005spc}, where it condenses on the nanocavity surface and red-shifts the wavelengths of the nanocavity modes.  In Fig.\ \ref{fig:tuning}(c), the PL spectra in the vicinity of the NV$^-$ ZPL are shown at discrete steps in the tuning process as the nanocavity mode closest to the NV$^-$ ZPL is tuned from 636~nm to 637.5~nm.  The nanocavity doublet resonance shifts diagonally across Fig.\ \ref{fig:tuning}(c) as the number of Xe tuning cycles increases.  ZPL emission between 636.8~nm-637.4~nm from several distinct NVs is visible, creating horizontal features whose center wavelength is unaffected by the Xe tuning. The inhomogeneous distribution of ZPL wavelengths is possibly the result of residual strain from the implantation step, or the diamond etching.  When the nanocavity modes cross the ZPL lines, in some cases the ZPL PL intensity increases.  In particular, a large enhancement is visible when the $\lambda_-$ nanocavity mode is resonant with a ZPL at 637.0~nm (labeled NV1). A comparatively modest enhancement is observed when $\lambda_-$ is resonant with a ZPL at 637.25~nm (labeled NV2). The relative magnitude of enhancement is shown in  Fig.\ \ref{fig:tuning}(d), which compares the PL spectrum when the nanocavity mode is on resonance with the NV1 ZPL (slice A in Fig.\ \ref{fig:tuning}(c)), on resonance with the NV2 ZPL (slice B in Fig.\ \ref{fig:tuning}(c)) and when it is detuned (slice C in Fig.\ \ref{fig:tuning}(c)).    

In general, the magnitude of the enhancement depends on the NV position and dipole orientation relative to the nanocavity field maximum, as indicated in Eq.\ \ref{eq:F}.    The nanocavity standing wave whispering gallery modes have well defined electric field amplitude nodes and anti-nodes, whose specific phase is determined by the nanocavity imperfections \cite{ref:borselli2005brs}. For maximum NV-cavity coupling, the NV must be located at an anti-node of one of the two standing wave modes, which is necessarily a node of the orthogonal standing wave mode.  This nanoscale sensitivity of the coupling strength on NV position is illustrated in Fig.\ \ref{fig:tuning}(c) by the asymmetry in the enhancement  of the NV1 ZPL enhancement when the nanocavity mode doublet peaks are on resonance.  The $\lambda_-$ mode strongly enhances the NV1 ZPL PL intensity; the $\lambda_+$ mode does not significantly affect the NV1 ZPL.  This indicates that NV1 is positioned near an anti-node of the $\lambda_-$ nanocavity mode.

To quantitatively determine the degree of enhancement, we measure the effect of the nanocavity mode on the spontaneous emission lifetimes of NV1 and NV2.  Figure \ref{fig:lifetime}(a) shows time resolved PL when the device is excited with a pulsed green source (4.75MHz repetition rate, 520~nm center wavelength, 28~nm bandwidth, 300~$\mu$W average power), with the nanocavity $\lambda_-$ mode is tuned on and off of resonance with NV1.  To ensure that the spontaneous emission properties of only the NV of interest were being measured, the PL was spectrally filtered using a monochromator centered at the wavelength of the NV1 ZPL (637.0~nm).  This emission was then directed to a time-correlated single photon counting module which records the photon detection time relative to the excitation pulse.    

Figures \ref{fig:lifetime}(b) and (c) show the spectra obtained for the on and off resonance measurements of NV1 under CW and pulsed excitation, respectively.  In both cases an enhancement to the NV1 ZPL intensity is observed when the nanocavity is on resonance.  Non-NV related emission from the nanocavity modes is strong in the pulsed excitation spectra (Fig.\ \ref{fig:lifetime}(c)), due to emission related to the nanocavity material or fabrication residue.  This emission decays quickly ($\sim3$~ns) compared to the NV emission, as shown by the inset to Fig.\ \ref{fig:lifetime}(a).  Figure \ref{fig:lifetime}(c) also reveals that the nanocavity $Q$ and doublet splitting was modified compared to the measurements in Fig.\ \ref{fig:tuning}.  During the lifetime measurements, $Q\sim3000$ and $\Delta\lambda\sim0.22$~nm, as determined from the fit in Fig.\ \ref{fig:lifetime}(c).  This degradation in $Q$, compared to the value observed during the measurements in Fig.\ \ref{fig:tuning}, may be related to repeated condensation and evaporation of Xe or other cyrostat contaminants over the course of several measurement cycles for this device.  Local modification of the GaP properties by the green excitation laser, as reported in Ref.\ \cite{ref:vandersar2011dnc}, may also affect the device characteristics.  

The NV1 spontaneous emission rate can be determined by fitting the data in  Fig.\ \ref{fig:lifetime}(a)  with simple exponential decay functions.  These fits indicate that the NV1 spontaneous emission lifetime is $\tau_{\text{c}} = 9.7\pm0.07$~ns and $\tau_\text{o} =11.6\pm0.3$~ns when  the nanocavity mode is on and off resonance, respectively, with the NV1 ZPL.  This indicates that the nanocavity enhancement of the ZPL emission appreciably modifies the total spontaneous emission rate of the NV.   From $\gamma_\text{o,c} = 1/\tau_{o,c}$, and the lifetime fits extracted from Fig.\ \ref{fig:lifetime}, we estimate $F_\text{ZPL}\sim6.3\pm 1.0$.  This corresponds to enhancing the  fraction of NV1 emission into the ZPL from $\zeta_\text{ZPL} \sim 3\%$ to $\sim 16\%$. 

The effect of the nanocavity on the NV2 ZPL is illustrated in Fig.\ \ref{fig:lifetime}(d-f), which  shows lifetime and spectral data when the nanocavity $\lambda_-$ mode is tuned on and off resonance.  These measurements were performed in the same manner as the NV1 measurements, but with the monochromator centered at the NV2 ZPL (637.2nm).   Note that during these measurements, the excitation spot alignment was optimized to maximize the NV2 ZPL emission, resulting in larger on and off resonance contrast in Fig.\ \ref{fig:lifetime}(e) compared to Fig.\ \ref{fig:tuning}(c).  Fits to the data in Fig.\ \ref{fig:lifetime}(d) indicate that the NV2 spontaneous emission lifetime is $\tau_\text{c} = 9.84\pm0.08$~ns and $\tau_\text{o} =11.0\pm0.2$~ns when  the nanocavity mode in on and off resonance, respectively, with the NV2 ZPL.  This indicates that $F_\text{ZPL} \sim 3.8 \pm 0.7$ for NV2.

From Eq.\ \ref{eq:F}, the FDTD predicted mode volume and field distribution presented above, and $Q\sim3000$ measured during the lifetime measurement, the maximum expected $F_\text{ZPL}$ can be calculated. For an NV optimally positioned relative to the nanocavity studied here, the ZPL emission would be enhanced by $F^\text{0nm}_\text{ZPL} = 12$ and $F_\text{ZPL}^\text{20nm} = 6.7$, assuming the NV is positioned at the surface, and 20nm below the surface (the maximum expected implantation depth), respectively.  $F_\text{ZPL}$ becomes smaller with the NV distance to the surface due to the evanescent decay of the nanocavity mode electric field intensity, which scales  roughly as $\exp(-2\kappa_z z)$, where $\kappa_z = (2\pi/\lambda)\sqrt{n_\text{GaP}^2 - n_\text{dia}^2}$.  $F_\text{ZPL}$ is also reduced when the NV is not positioned at an anti-node of the standing-wave mode, and if the NV dipole is misaligned relative to the field polarization.   As discussed above, the relative degree of ZPL enhancement when it is resonant with the $\lambda_-$  and $\lambda_+$ modes in Fig.\ \ref{fig:tuning}(a) suggests that NV1 is closely positioned near a node of the $\lambda_+$ standing wave, and an anti-node of the $\lambda_-$ standing wave. 

In future work, $F_\text{ZPL}$ can be increased using higher $Q$ nanocavities, and by further reducing $V$. Additional studies are required to determine the mechanism limiting the $Q$ of the devices studied here, as well as the degradation in $Q$ observed over the course of the measurements.  Surface roughness and other defects are visible in the SEM image in Fig.\ \ref{fig:sem}(a), and the large mode splitting $\Delta\lambda$ indicates that surface scattering is present in these devices and likely plays an important role in limiting $Q$.  By eliminating this roughness, it should be possible to achieve $Q$ in excess of $2\times 10^4$, as was demonstrated in larger diameter hybrid GaP-diamond devices\cite{ref:barclay2009cbm}.  Achieving this $Q$ with the specific coupled NV-nanocavity device studied here would increase $F_\text{ZPL}$ to $\sim 42$, corresponding to enhancing the ZPL branching ratio to $\zeta_\text{ZPL} \sim 57\%$.

Shrinking $V$ is immediately possible by coupling an NV to the TM mode ($m=10$, $p=0$) of the nanocavity studied here. The TM mode has a smaller mode volume, $V\sim 2.6(\lambda/n_\text{GaP})^3$, than the TE mode, and a larger relative intensity at the diamond surface, $\eta_\text{dia}\sim0.15$ , resulting in an approximately $52\%$ increase in $F^\text{0nm}_\text{ZPL}$ compared to the TE mode for a given $Q$ and an optimally positioned NV.  Further reduction in $V$ is possible by decreasing the nanocavity diameter to $d=650$~nm; the resulting device supports a TM mode with $V\sim 1.7(\lambda/n_\text{GaP})^3$ and $\eta_\text{dia}\sim0.39$ , and radiation $Q_\text{rad}\sim4\times10^4$.  Ultrasmall $V$ is possible using a hybrid GaP-diamond photonic crystal nanocavity proposed in Ref.\ \cite{ref:barclay2009hpc}, which supports TE modes with $V\sim0.5(\lambda/n_\text{GaP})^3$ while maintaining a radiation limited $Q_\text{rad}>10^6$. These photonic crystal nanocavities are naturally suited to being interconnected, and are a promising system for implementing an on-chip quantum network to enable interactions between NVs.  

In the immediate future, ZPL photon detection rates exceeding those achievable using a microscope objective to collect NV emission can be achieved by combining the Purcell enhancement from the ring nanocavites demonstrated here,  and the fiber taper waveguide PL collection studied in Ref.\ \cite{ref:fu2011ltt}.  This would be an important resource for quantum information resources such as NV-photon entanglement\cite{ref:togan2010qeb} which rely on narrowband measurement of ZPL emission. More generally, a fiber coupled nanocavity-NV system could be used as a bright narrowband source of indistinguishable photons, and to improve readout rates of the NV spin properties. 


This material is based upon work supported by the Defense Advanced Research Projects Agency under Award No. HR0011-09-1-0006 and The Regents of the University of California.


\end{document}